\documentstyle[eqsecnum,multicol,aps,preprint]{revtex}
\begin{document}
\preprint {}
\title {Nonequilibrium phase transition in the kinetic Ising model: Existence of tricritical
point and Stochastic resonance}
\author {Muktish Acharyya$^+$}
\address{Theoretische Physik, Gerhard-Mercator-Universit\"at Duisburg, 47048 Duisburg, Germany}
\date{\today}
\maketitle
\begin{abstract}
The dynamic phase transition has been studied in the 
two dimensional kinetic Ising model in presence of
a time varying (sinusoidal) magnetic field by Monte Carlo simulation. 
The nature (continuous or discontinuous) of the transition
is characterized by studying the distribution of the order parameter and the temperature variation of
the fourth order cumulant. For the higher values of the field amplitude the transition observed
is discontinuous and it is continuous for lower values of the field amplitude, indicating the existence
of a {\it tricritical point} (separating the nature of transition) 
on the phase boundary. The transition is observed to be a manifestation of {\it stochastic resonance}. 

\bigskip
\bigskip
\bigskip

\begin{flushleft}{{\bf PACS: 05.50.+q}}\end{flushleft}
\end{abstract}

\newpage

\section{Introduction}

The kinetic Ising model in presence of an oscillating magnetic field gives rise to various interesting
dynamical responses \cite{rik1}. 
The dynamic phase transition, hysteresis \cite{ac} and {\it stochastic resonance}
\cite{sr} are most important
dynamic responses of recent interest. 
Tome and Oliveira \cite{tom} first 
observed and studied the dynamic transition in the
kinetic Ising model in presence of a sinusoidally oscillating magnetic field. 
They solved the mean field (MF)
dynamic equation of motion (for the average magnetization) of the kinetic
Ising model in presence of a sinusoidally oscillating magnetic field.
By defining the order parameter as the time averaged
magnetization over a full cycle of the 
oscillating magnetic field they showed that 
the order parameter vanishes
depending upon the value of the temperature and the 
amplitude of the oscillating field. 
In the field amplitude and temperature plane they have drawn
a phase boundary separating dynamic ordered 
(nonzero value of the order parameter) and disordered (order 
parameter vanishes) phases. They \cite{tom} have also predicted
a {\it tricritical point} (TCP),
separating the nature (discontinuous/continuous) of the transition
on the phase boundary line. 
However, such a transition, observed \cite{tom} from the solution of mean field
dynamical equation, is not dynamic in true sense.
This is because, for the field amplitude less than the
coercive field (at temperature less than the transition temperature
without any field), the response magnetization varies periodically
but asymmetrically even in the zero frequency limit; the system remains locked
to one well of the free energy and cannot go to the other one, in the absence
of noise or fluctuation.

Lo and Pelcovits \cite{lo} first attempted to study the dynamic 
nature of this phase transition (incorporating the effect of fluctuation)
in the kinetic Ising model by Monte Carlo (MC) simulation. 
In this case, the transition disappears in the 
zero frequency limit; due to the 
presence of fluctuations, the magnetization flips to the
direction of the magnetic field and the dynamic order parameter 
vanishes.
However, they \cite{lo} have 
not reported any precise phase boundary.
Acharyya and Chakrabarti \cite{ac} studied the nonequilibrium dynamic phase
transition in the kinetic Ising model 
in presence of oscillating magnetic field by 
extensive MC simulation. They \cite{ac} have successfully drawn the phase
boundary for the dynamic transition and predicted a tricritical 
point on it.
It was also noticed by them \cite{ac}
that this dynamic phase transition is associated with the 
breaking of the symmetry of
the dynamic hysteresis loop. 
In the dynamically disordered (value of order parameter vanishes)
phase the corresponding hysteresis loop is
symmetric, and loses its symmetry in the ordered phase (giving
nonzero value of dynamic order parameter).

Recent studies which reveal the thermodynamic nature of the dynamic transition are 
on the temperature variations of ac susceptibility \cite{ac},
on the relaxation behavior of dynamic order parameter and
divergence of time scale (critical slowing down) \cite{ma1},
on the scaling
of the distribution of dynamic order parameter and the divergence of length scale 
\cite{rik2} and on the temperature variation
of the dynamic correlation \cite{ma2}. 

Although the existence of a
TCP has been predicted from the temperature variations of the
average order parameter \cite{ac,tom}, a detailed and systematic study has
not yet been performed to detect the nature (continuous/discontinuous) of the dynamic transition 
along the dynamic phase boundary.
In this paper, the statistical distribution of the dynamic order parameter has been studied
to detect the nature of the transition, by Monte Carlo simulation in a two dimensional
kinetic Ising model in presence of an oscillating magnetic field. The temperature variation
of the fourth order cumulant \cite{binder} (of the distribution of dynamic order parameter)
has also been studied to characterize the transition.
The relation between stochastic resonance \cite{sr} and dynamic transition \cite{ac} is
also discussed.
The paper
is organized as follows: in section II the model and the MC simulation scheme are discussed,
section III contains the simulational results,
the paper ends
with a summary of the work in section IV.

\section{Description of the Model and the simulation scheme}

The Hamiltonian, of an Ising model (with ferromagnetic nearest
neighbor interaction) in presence of a time varying magnetic field, can
be written as
\begin{equation}
H = -J\sum_{<ij>} s_i s_j - h(t) \sum_i s_i.
\label{hm} 
\end{equation}
Here, $s_i (=\pm 1)$ is the Ising spin variable, $J > 0$ is the ferromagnetic spin-spin interaction
strength and
$h(t)$  
is the sinusoidally oscillating (in time but uniform in space) magnetic field. The
time variation of $h(t)$ can be expressed as

\begin{equation}
h(t) = h_0 \cos (\omega t)
\label{field}
\end{equation}
\noindent where $h_0$ is the amplitude and $\omega (= 2\pi f)$ is the angular 
frequency of the oscillating field.
The system
is in contact with an isothermal heat bath at temperature $T$. 

A square lattice (with periodic boundary condition) of linear size $L (=100)$ is considered.
The initial condition is that randomly $50\%$ of all spins are up (+1).
At any finite
temperature $T$,
the dynamics of this system has been studied here by Monte Carlo
simulation using Metropolis single spin-flip dynamics \cite{binder}. The transition rate 
is specified as
\begin{equation}
W(s_i \to -s_i) = {\rm Min}\left[1, \exp(-{\Delta H}/K_BT)\right]
\end{equation}
\noindent where $\Delta H$ is the change in energy due to spin flip ($s_i \to -s_i$)
 and $K_B$ is the Boltzmann
constant.
Any lattice site is chosen randomly and the spin variable ($s_i^z$) is updated according to the
Metropolis probability.
$L^2$ such updates constitute
the time unit (Monte Carlo step per spin or MCSS)
here. The magnitude of the field $h(t)$ changes after every MCSS following
equation \ref{field}. 
The instantaneous magnetization
(per site),
 $m(t) = (1/L^2) \sum_i s_i^z $ has been calculated. 

The time averaged (over the complete cycle of the oscillating magnetic field)
magnetization $Q = {1 \over {\tau}}
\oint m(t) dt$ defines the dynamic order parameter \cite{tom}. 
The frequency is $f = 0.001$ (kept fixed throughout the study). So, one complete cycle of the oscillating
field takes 1000 MCSS (time period $\tau$ = 1000 MCSS). 
A time series of magnetization $m(t)$ has been generated up to $10^6$ MCSS. This time series contains
$10^3$ (since $\tau$ = 1000 MCSS) 
number of cycles of the oscillating field. The dynamic order parameter $Q$ has been calculated
for each such cycle.
So, the statistics 
(distribution of $Q$) is based on $N_s = 10^3$ different
values of $Q$. 
The fourth order cumulant \cite{binder} (dynamic order parameter) is defined as
\begin{equation}
U_L = 1.0 - <Q^4>/(3<Q^2>^2)
\label{cum}
\end{equation}
where, $<Q^n> = \int Q^n P(Q) dQ$ and $P(Q)$ is the normalized ($\int P(Q) dQ = 1$) distribution of $Q$.

The computational speed recorded is 1.42 MUPS (Million Updates Per Second) in RS6000/43p of IBM cluster.

\section{Results}

The statistical distribution $P(Q)$ of dynamic order parameter $Q$
and its temperature dependence
have been studied closed to the phase boundary to detect the nature 
of the transition.
Figure 1 shows the distributions (at fixed
value of the field amplitude) for three different values of temperature. Below, the transition (Fig. 1a)
the distribution shows only two equivalent peaks
centered around $\pm 1$. Close to the transition point (Fig. 1b), a third peak
centered around zero is developed. As the temperature increases slightly (Fig. 1c), the strength of the
third peak increases in cost of that of two other (equivalent) peaks.
Above the transition (Fig. 1d), only one peak is observed centered around zero. 
This indicates \cite{binder} that the
transition is first order or discontinuous.

What is the origin of this kind of first order transition ? To get the answer of this question,
the time variation of the magnetization $m(t)$ is studied (in Fig. 2) for several cycles of the oscillating
magnetic field $h(t)$, close to the transition. 
From fig. 2 it is clear that sometimes, the system likes to 
stay in the positive well (of the double well form of the free energy) 
and sometimes it likes to stay in other. It
is obvious that the best time for the system to switch from one well to the other one, is
when the value of the field is optimum ("good opportunity") \cite{sr}.
So, if the system misses one "good opportunity" (first half period of the oscillating field)
to jump to the other well it has to wait for a new chance (another full period of the oscillating field).
Consequently, it shows that the residence time (staying time in a particular well) can only be 
nearly equal to an odd
integer multiple of the half-period (half of the time period of the oscillating field)
\cite{sr}. This leads to two consequences: 

(1) The distribution of the 
dynamic order parameter $Q$ would be peaked around three values (i) $Q \approx 0$, when the system 
utilizes "good opportunity" and goes from one well to the other (marked 'A' in
Fig. 2), (ii) $Q \approx -1$, when the
system misses the "good opportunity" to go from negative well to the 
positive well and it stays for one (or
more) full period in the negative well (marked 'B' in Fig. 2), 
(iii) $Q \approx +1$, when the system misses the "good opportunity" to
go from positive well to negative well and spends one (or more) full period
in the positive well (marked 'C' in Fig. 2). 
As a result, the distribution of $Q$ would give three distinct peaks centered
at +1, -1 and 0.

(2) The other consequence of this kind of time variation, of magnetization $m(t)$, is
the "stochastic resonance". 
This can be detected from the distribution of residence time (the time system spends in a particular
well).
From Fig. 2 it is 
clear that, the distribution ($P_r$) of residence time ($\tau_r$) 
will be peaked multiply around the odd integer multiple of
half-period
\cite{sr}. One such distribution is shown in Fig. 3. The distribution shows multiple peaks around the
odd integer values (500, 1500, 2500, 3500, 4500 and 5500 MCSS)
of half-period ($\tau/2$=500 MCSS,of the driving fields). The heights of the peaks decreases
exponentially (dotted line in Fig. 3) with the peak positions. This is the fingureprint
of stochastic resonance \cite{sr}.

The fourth order cumulant \cite{binder}
$U_L$ has been plotted against the temperature.
In the case of discontinuous transition, the simultaneous appearance of three peaks (of the
distribution of dynamic order parameter), is responsible for very high value of $<Q^4>$ (compared
to the value of $3<Q^2>^2$) at the transition point. This will lead to a deep minimum (with large
negative value) of fourth
order cumulant $U_L$ at the transition point.
So, the deep minimum corresponds to the first order transition and the position of minimum 
is related to the transition point (Fig.4).
From the above observations it is clear that the transition (across the upper part of the
dynamic phase boundary) is first order and a manifestation of stochastic resonance.

Figure 5 shows the distributions of the dynamic order parameter $Q$
for three different values of the temperature. Here, the field amplitude $h_0$ is quite low in
comparison with that used
in the earlier case (Fig. 1).
It shows that, in the ordered region, this gives two (equivalent)
peaks (Fig. 5(a)) and as the temperature increases these two peaks come close to each other continuously 
(Fig. 5(b)) and close 
to the transition (and also above it) (Fig. 5(c))
only one peak (centered around zero) is observed. This feature reveals the continuous
or second order transition \cite{binder}. 
The second order transition is also characterized by the temperature variation of the
fourth order cumulant $U_L$ Eq.~(\ref{cum}). 
Figure 6 shows that $U_L$ continuously decreases from 2/3 to zero
revealing the second order phase transition
\cite{binder}. It should be mentioned here that the 
temperature variation of the cumulant and the
finite size study (in the continuous transition region) has
been made by Sides et al \cite{rik2}, here it has been reexamined for completeness. It is important
to note that they \cite{rik2} studied the dynamic transition by varying the frequency (keeping the
temperature and field amplitude fixed), whereas, the present study has been done by varying the
temperature (fixing frequency and amplitude of the field). However, it is believed that the results are
qualitatively invariant under the choice of tunable parameter.

\bigskip
\section{Summary}

The nonequilibrium dynamic phase transition has been studied in the kinetic Ising model in presence of
a time varying (sinusoidal) magnetic field by Monte Carlo simulation. The nature of the transition
is characterized by studying the distribution of order parameter and the temperature variation of
fourth order cumulant. For the higher values of the field amplitude the transition observed
is discontinuous and it is continuous for lower values of the field amplitude. This indicates that
there is a tricritical point [separating the nature (continuous/discontinuous) of the
dynamic transition] located on the dynamic phase boundary. 
These observations supports the earlier predictions \cite{ac,tom} of a TCP on the phase boundary.
The residence time distribution shows that the transition is a manifestation of {\it stochastic
resonance}. 
A lengthy computational effort is required 
to find the precise location of the tricritical point. It would be interesting to know whether
the TCP can act as a limit of stochastic resonance (along the first order line) or not. 
An extensive investigation is going on towards this direction 
and the results will be reported elsewhere.

The detailed finite size study has been performed by Sides et al \cite{rik2} and 
they have not observed any discontinuous transition. They studied the dynamic
transition in very high frequency range. For very high frequency, the tricritical point will shift
towards the zero temperature \cite{tcp}
(the region of first order transition on the phase boundary will be very short). For this reason
Sides et al. overlooked the part of dynamic phase boundary corresponding to first order
transition.
The first order region of the dynamic phase boundary can be observed clearly in low
frequency range.

The experimental evidence \cite{jiang} of the dynamic transition has been found recently.
Dynamical symmetry breaking (associated with the dynamic transition) of the hysteresis loop
across the transition point has been observed in highly anisotropic (Ising like) and 
ultra thin Co/Cu(001) ferromagnetic films by the surface magneto optic Kerr effect. Dynamical
symmetry breaking of the hysteresis loop has also been observed \cite{suen} in ultra thin
Fe/W(110) films. However, the detailed investigation has not yet been made to study the
dynamic phase boundary and the nature (continuous/discontinuous) of the transition.

\section*{Acknowledgments}
Graduiertenkolleg for "Structure and Dynamics of heterogeneous systems"
is gratefully acknowledged for financial support. Author would like to thank B. K.
Chakrabarti, S. L\"ubeck, U. Nowak and K. Usadel for careful reading of the manuscript and comments.

\begin{figure}
\caption
{The histograms of the normalized distributions of the dynamic order parameter $Q$
for different temperatures ($T = 0.20J/K_B$, $0.28J/K_B$, $0.30J/K_B$ and $0.40J/K_B$) 
and for the fixed value of field amplitude $h_0$.
All the figures are plotted in the same scales.}
\label{probq1}
\end{figure}
\begin{figure}
\caption
{Time variation of the magnetic field $h(t)$ 
(solid line) and magnetization $m(t)$ (dotted
line) close to the transition ($T = 0.3J/K_B$ and $h_0 = 2.0J$).}
\label{mht}
\end{figure}

\begin{figure}
\caption
{The histogram of normalized ($\int P_r(\tau_r) d\tau_r = 1$)
distribution ($P_r(\tau_r)$) of the residence
time ($\tau_r$). 
The dotted line is the exponential best fit of the envelope
of the distribution.}
\label{sr}
\end{figure}
\begin{figure}
\caption
{Temperature ($T$) 
variation of the fourth order Binder cumulant. Deep minimum indicates the
transition is first order and the position of minimum is the transition point.}
\label{cum1}
\end{figure}

\begin{figure}
\caption
{The normalized distributions of the dynamic order parameter $Q$ ( in the 2nd order
and close to the transition region ) for three different temperatures ($T = 1.48J/K_B$, 
$1.50J/K_B$, $1.55J/K_B$) and fixed field amplitude
$h_0 = 0.3J$.} 
\label{probq2}
\end{figure}
%
\begin{figure}
\caption
{Temperature ($T$) variation of the fourth order cumulant ($U_L$) for
a fixed value of field amplitude ($h_0 = 0.3J$).}
\label{cum2}
\end{figure}


\begin{references}
\bibitem{tcp} 
The tricritical point on the phase boundary appears because of the failure to relax within
the time period ($2\pi/\omega$). This intrinsic relaxation time in the
ferromagnetic phase decreases as the temperature decreases and below $T_{TCP}(h_0, \omega)$, the
effective relaxation time ($\tau_{eff}$) is less than $2\pi/\omega$. The TCP will be located at the
temperature where $\tau_{eff} \approx 2\pi/\omega$. This indicates that TCP will shift towards low
temperature as the frequency increases. For detailed discussion see Ref. \cite{ac}.
\end{references}
\end{document}